\documentclass[twocolumn,aps,prb]{revtex4}
\usepackage{graphicx}
\usepackage{amssymb}
\usepackage{amsmath}
\usepackage{bm}
\usepackage{color}
\def\Journal #1,#2,#3,#4#5#6#7{#1 {\bf #2}, #3 (#4#5#6#7)}

\def\s{\sigma}
\def\Vec#1{{\bf #1}}

\begin{document}

\title{Magnetotransport in Weyl semimetal nanowires}
\author{Akira Igarashi}
\affiliation{Department of Physics, Tohoku University, Sendai 980-8578, Japan}
\author{Mikito Koshino}
\affiliation{Department of Physics, Osaka University, Toyonaka 560-0043, Japan}
\date{\today}

\begin{abstract}
We theoretically study the band structure and the electronic transport in the Weyl semimetal nanowires
in magnetic fields, and demonstrate that the interplay of the Fermi-arc surface states and the bulk Landau levels
plays a crucial role in the magnetotransport.
We show that a magnetic field perpendicular to the surface
immediately hybridizes the counter-propagating surface modes
into a series of dispersionless 0th Landau levels,
and it leads to a significant reduction of the traveling modes and a rapid decay of the conductance.
On the contrary, a magnetic field parallel to the wire 
adds linearly-dispersed 0th  Landau levels to the traveling modes and 
increases the conductance.
\end{abstract}



\maketitle

\section{Introduction}

A Weyl semimetal is a three dimensional electronic system in condensed matter,
which hosts the massless chiral Fermions in the low-energy excitation.
\cite{murakami2007phase, burkov2011weyl, burkov2011topological, wan2011topological, yang2011quantum,hosur2013recent}
The energy spectrum of the Weyl semimetal 
is characterized by topologically-protected band touching points called the Weyl nodes,
around which the energy bands linearly disperse in all three directions.
One of the most distinctive features of the system is 
the coexistence of the topological surface states and the bulk massless Fermion states.
The Fermi surface of the surface states always takes a form 
of a finite segment terminated at the Weyl nodes, and this is called Fermi arc.\cite{wan2011topological}
The existence of the Fermi arc was experimentally verified in several materials by the photoemission spectroscopy.
\cite{xu2015observation,weng2015weyl,lu2015experimental,xu2015discovery,lv2015experimental,huang2015weyl,xu2015experimental}

It is intriguing to consider how the Fermi-arc surface states contribute to the electronic transport.
The exotic transport properties of the Weyl semimetal intensively studied in the previous works, 
represented by the chiral anomaly effect\cite{nielsen1983adler,aji2012adler,son2013chiral,burkov2014chiral,gorbar2014chiral,lu2015high,burkov2015negative,kim2013dirac,li2016negative,li2015giant,xiong2015evidence,PhysRevX.5.031023,
yang2015chiral,zhang2016signatures}, are understood as intrinsic properties of the relativistic Weyl Fermions in the bulk states.
On the other hand, several works have focused on the direct effects of the surface states 
on the various transport phenomena. \cite{potter2014quantum,gorbar2014quantum,bulmash2016quantum,zhang2016quantum,haldane2014attachment,parameswaran2014probing,baum2015current,ominato2016magnetotransport,baireuther2016scattering,gorbar2016origin,inoue2016quasiparticle,moll2016transport,siu2016dirac} 
It was predicted that an open Fermi arc gives rise to anomalous quantum oscillation 
in Weyl semimetal thin film \cite{potter2014quantum,gorbar2014quantum,bulmash2016quantum,zhang2016quantum},
and it was observed in a recent experiment for the Dirac semimetal Cd$_3$As$_2$, \cite{moll2016transport}
which has doubly degenerate Weyl-semimetal bands overlapping in the momentum space.
Concerning other topological systems, the surface-state transport was studied for the nanowire of topological insulators, where the helical surface states give rise to the Aharanov-Bohm oscillation with non-trivial Berry phase.
\cite{bardarson2010aharonov,zhang2010anomalous,hong2014one,cho2015aharonov,jauregui2016magnetic}
Similar magnetic oscillation was also observed in the Dirac semimetal. \cite{wang2016aharonov}

\begin{figure}
\begin{center}
\leavevmode\includegraphics[width=1.\hsize]{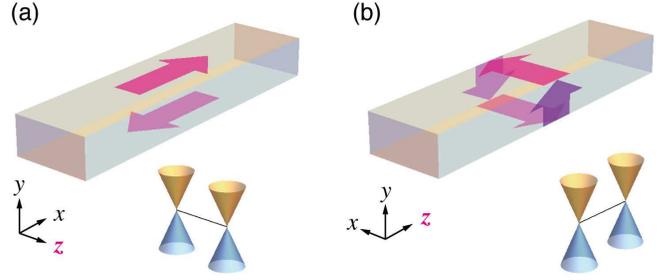}
\end{center}
\caption{Schematics for the Weyl-semimetal wires
which are (a) perpendicular and (b) parallel to the Weyl-node splitting direction ($z$ direction).
Arrows indicate the propagation of the surface states.
The case (a) is considered in this paper.
}
\label{geo_tube}
\end{figure}

In this paper, we theoretically study the magnetotransport of Weyl semimetal nanowires,
and demonstrate that the interplay of the surface states and the bulk Landau levels,
which is a hallmark of the Weyl semimetal, makes a significant impact on the electronic conductance.
Here we model a nanowire using a tight-binding lattice model, 
and consider a situation illustrated by Fig.\ \ref{geo_tube}(a), 
where the topological surface states travel along the wire.
We then calculate the band structure and the electric conductance under various magnetic field directions.
We find that $B$-field perpendicular to the surface states ($B \parallel y$)
immediately hybridizes the counter-propagating surface modes on the opposite faces,
leading to a rapid decay of the conductance.
In increasing the magnetics field, the surface linear bands eventually cross over 
to a series of dispersionless $n=0$ Landau levels,
causing a quantum oscillation 
which is consistent with the slab geometry. \cite{potter2014quantum}
If $B$-field is parallel to the wire  ($B \parallel x$), on the other hand,
the dispersive $n=0$ Landau levels are added to conducting channels and the conductance increases.

\section{Model and formulation}
To describe the Weyl semimetal,
we consider a 2 by 2  cubic-lattice Hamiltonian given by 
\begin{align}
H = &2 t [\sigma_x\sin k_x a + \sigma_y\sin k_y a + \sigma_z \cos k_z a  \nonumber \\
   & \qquad + \sigma_z (2-\cos k_x a-\cos k_y a)]
\label{eq_H}
\end{align}
 where $\sigma_i\, (i=x,y,z)$ are the Pauli matrices, $a$ is the lattice constant, 
 and $t$ is the parameter for energy scale.
 The corresponding Schr\"odinger equation in the real space becomes
\begin{align}
&(E/t) \psi(\Vec{r}) = \notag\\
& \quad + (-i\s_x-\s_z)\psi(\Vec{r}+\Vec{a}_x) + (i\s_x-\s_z)\psi(\Vec{r}-\Vec{a}_x) \notag \\ 
& \quad + (-i\s_y-\s_z)\psi(\Vec{r}+\Vec{a}_y) + (i\s_y-\s_z)\psi(\Vec{r}-\Vec{a}_y) \notag \\ 
& \quad + \s_z\psi(\Vec{r}+\Vec{a}_z) + \s_z\psi(\Vec{r}-\Vec{a}_z)  + 4\s_z \psi(\Vec{r}),
\label{eq_sch}
\end{align}
where $\Vec{a}_i = a \, \Vec{e}_i$, $\Vec{e}_i$ is the unit vector in $i(=x,y,z)$ direction, 
and $\psi(\Vec{r})$ is the two-component wave function at the lattice point $\Vec{r}$.
The energy eigenvalues of Eq.\ (\ref{eq_H}) are
 \begin{align}
 E = &\pm 2t \Bigl\{
 \sin ^2 {k_x a} + \sin ^2 {k_y a} 
\notag\\ 
& \hspace{-1mm} + \left[\cos k_z a + (2-\cos k_x a-\cos k_y a) \right]^2
 \Bigr\}^{1/2}.
\label{eq_E}
\end{align}
The conduction and valence bands are touching at two Weyl nodes at 
$\Vec{k}_\pm=(0,0,\pm \pi/(2a))$.
The Hamiltonian near $\Vec{k}_\pm$ is approximated within the linear order by
the Weyl Hamiltonian,
\begin{align}
 H^{\rm (eff)}_{\pm } = \hbar v(k_x \sigma_x +k_y \sigma_y \pm k_z \sigma_z),
 \label{eq_H_eff}
\end{align}
where $\hbar v = 2ta$, and $(k_x,k_y,k_z)$ is the relative wave vector measured from $\Vec{k}_\pm$.

When we apply a magnetic field $\Vec{B}$ to the system,
the matrix element from the site $\Vec{r}_i$ to $\Vec{r}_j$ in Eq.\ (\ref{eq_sch})
acquires the Peierls phase factor $e^{- i \phi_{ij}}$
\begin{align}
\phi_{ij} = -\frac{e}{\hbar}\int_{\Vec{r}_i}^{\Vec{r}_j} \Vec{A}(\Vec{r}) \cdot d  \Vec{r},
\end{align}
where $\Vec{A}(\Vec{r})$ is the vector potential giving $\Vec{B} = \nabla \times \Vec{A}$.
For a uniform $B$-field, $\Vec{B}=(B_x,B_y,B_z)$, we set the gauge as $\Vec{A} = (B_y z - B_z y,0,B_x y)$
independently of $x$, so that the eigen states can be labeled by the parallel momentum $k_x$ along the wire.
The magnetic-field amplitude is characterized by the flux per unit cell, 
\begin{equation}
\phi = \frac{Ba^2}{h/e}.
\end{equation}


We consider an infinitely long channel with rectangular cross section
using this lattice model.
Now Eq.\ (\ref{eq_H}) is anisotropic in that the two Weyl nodes are split in $z$ direction.
Accordingly, there are two distinct geometries as shown in Fig.\ \ref{geo_tube},
where the wire is  (a) perpendicular or (b) parallel to $z$ direction.
The surface states always appear on the faces parallel to $z$ axis, 
on which the surface electrons flow perpendicularly both to $z$ and the surface normal.
In this paper, we consider the situation (a), where the surface current is parallel to the wire,
to investigate the magnetic field effect on the surface state transport.
In this geometry, the wire stretches along $x$-direction, and has a rectangular cross section of $L_y\times L_z$.
The wave function of the surface state spreads on the faces perpendicular to the $y$-axis,
and the associated electronic current flows to $x$-axis with the opposite directions on the opposite faces.
The geometry (b) (not considered here) corresponds to the situation in the recent experiment for Cd$_3$As$_2$ microstructure. \cite{moll2016transport}

We calculate the two-terminal conductance of the wire 
by assuming the chemical potentials $\mu_1$ and $\mu_2\,(\mu_1 >\mu_2)$
at the left end ($x=-\infty$) and right end ($x=+\infty$)  of the wire, respectively.
We neglect impurity effects and relaxation 
between the left-going and the right-going channels by assuming the ballistic situation.
The electric current is then given by
\begin{align}
I = -e \int_{\mu_2}^{\mu_1} dE \sum_n \int \frac{dk_x}{2\pi}\,v_{n,k_x} \theta_{n,k_x} \delta(E-E_{n,k_x}),
\end{align}
where $E_{n,k_x}$ is the dispersion of the $n$-th energy bands of the wire,
$v_{n,k_x} = \partial E_{n,k_x}/ \hbar\partial k_x$ is the band velocity,
and $\theta_{n,k_x} =1$ and 0 for right-going states $(v_{n,k_x} > 0)$
and left going states $(v_{n,k_x} < 0)$, respectively.
Here we set $\mu_1-\mu_2$ to a finite but small value, $0.02t$,
to avoid the contribution of the nearly-flat bands having exponentially small band velocities.
The conductance is given by $G = I/(\mu_1-\mu_2)/(-e)$.

\begin{figure}
\begin{center}
\leavevmode\includegraphics[width=1.\hsize]{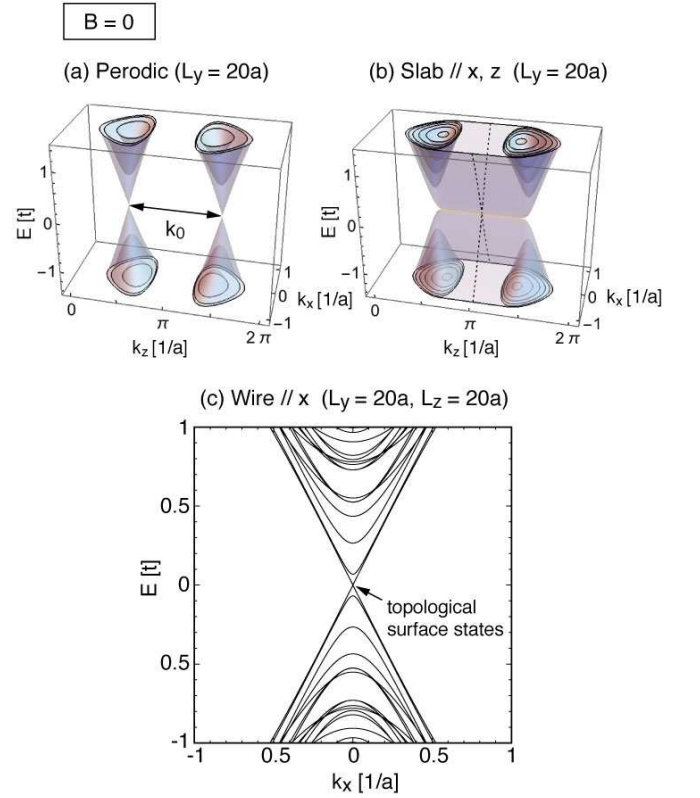}
\end{center}
\caption{
Band structures of the lattice model Eq.\ (\ref{eq_H}) with $B=0$,
for (a) periodic boundary condition with $L_y=20a$, 
(b)  slab geometry with thickness $L_y=20a$.
and (c) wire geometry with $L_y=L_z=20a$.
Linear bands crossing at $E=0$ are the surface-state bands.
}
 \label{band_b0}
\end{figure}

\section{Band structures and conductance}

\begin{figure*}
\begin{center}
\leavevmode\includegraphics[width=1.\hsize]{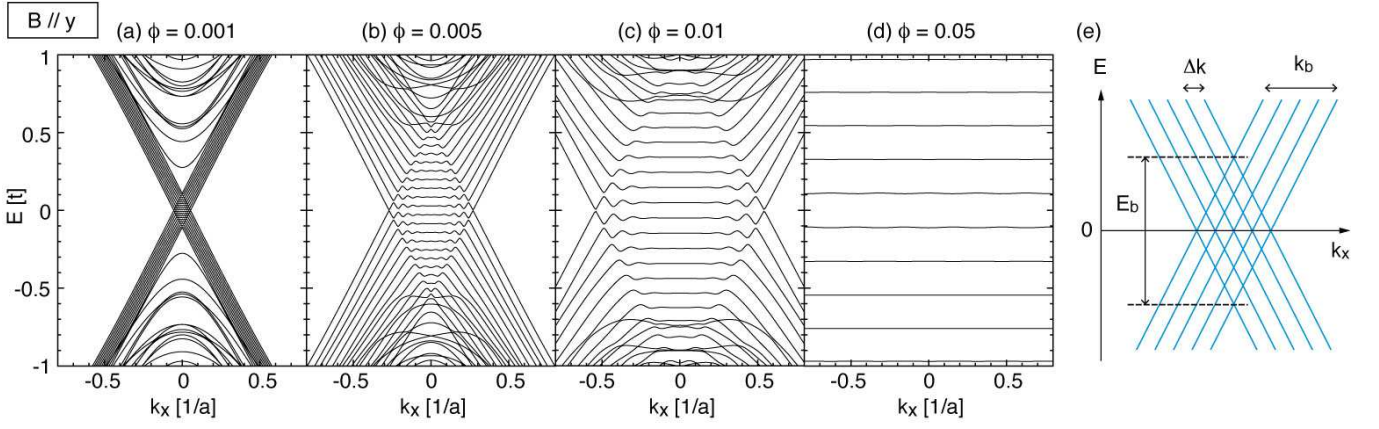}
\end{center}
\caption{(a-d) Band structures calculated for the wire of $L_y=L_z=20a$, 
in $B \parallel y$ with the amplitude of (a) $\phi =0.001$ (b) 0.005 (c) 0.01 and (d) 0.05.
(e)  Schematics for the surface band crossing in the weak field regime.
}
\label{band_magy}
\end{figure*}

\subsection{Zero magnetic field} 

In the finite-sized Weyl semimetal, the surface state band appears between
the Weyl nodes to make the Fermi arc.
Figures \ref{band_b0} present the band structures of the lattice model Eq.\ (\ref{eq_H}),
with (a) the periodic boundary condition in $y$ with period $L_y=20a$, 
and (b) the closed boundary condition (slab geometry)  in $y$ with thickness $L_y=20a$.
In the latter,  we see the surface bands spanning the region between the Weyl nodes,
and they linearly disperse in $k_x$ while almost independent of $k_z$.

If we take a wire geometry which is finite in both $y$ and $z$-direction,
the surface bands in Fig.\ \ref{band_b0}(b) are further discretized in $k_z$ direction
and form degenerate linear bands.
Figure \ref{band_b0}(c)  plots the energy bands of the wire with $L_y=L_z=20a$.
The linear bands crossing at $E=0$ are the degenerate surface states, 
where the positive and negative slopes are contributed from the surface states located 
at $y=L_y/2$ and $-L_y/2$, respectively.
The number of the degenerate bands in each slope is approximately written as
\begin{align}
N_0 \approx \frac{k_0}{2\pi/L_z},
\label{eq_N0}
\end{align}
where $k_0=\pi/a$ is the span between the Weyl nodes.
In the case of Fig.\ \ref{band_b0}(b), we have $N_0=9$.
As we have $N_0$ left-going channels and $N_0$ right-going channels,
the conductance in the ideal system is given by
\begin{align}
G = \frac{e^2}{h}N_0.
\label{eq_G0}
\end{align}

\begin{figure}
\begin{center}
\leavevmode\includegraphics[width=1.\hsize]{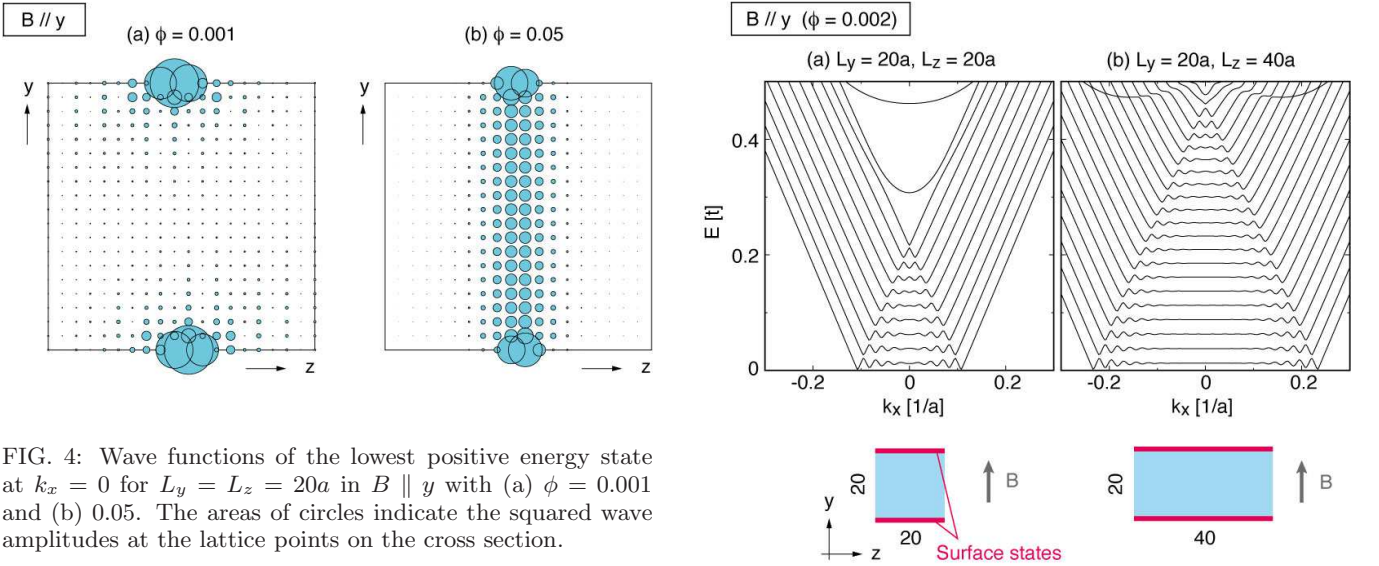}
\end{center}
\caption{Wave functions of the lowest positive energy state at $k_x=0$
for $L_y=L_z=20a$ in $B \parallel y$ with (a) $\phi =0.001$ and (b) 0.05.
The areas of circles indicate the squared wave amplitudes at the lattice points on the cross section.
}
 \label{fig_wave}
\end{figure}

\begin{figure}
\begin{center}
\leavevmode\includegraphics[width=1.\hsize]{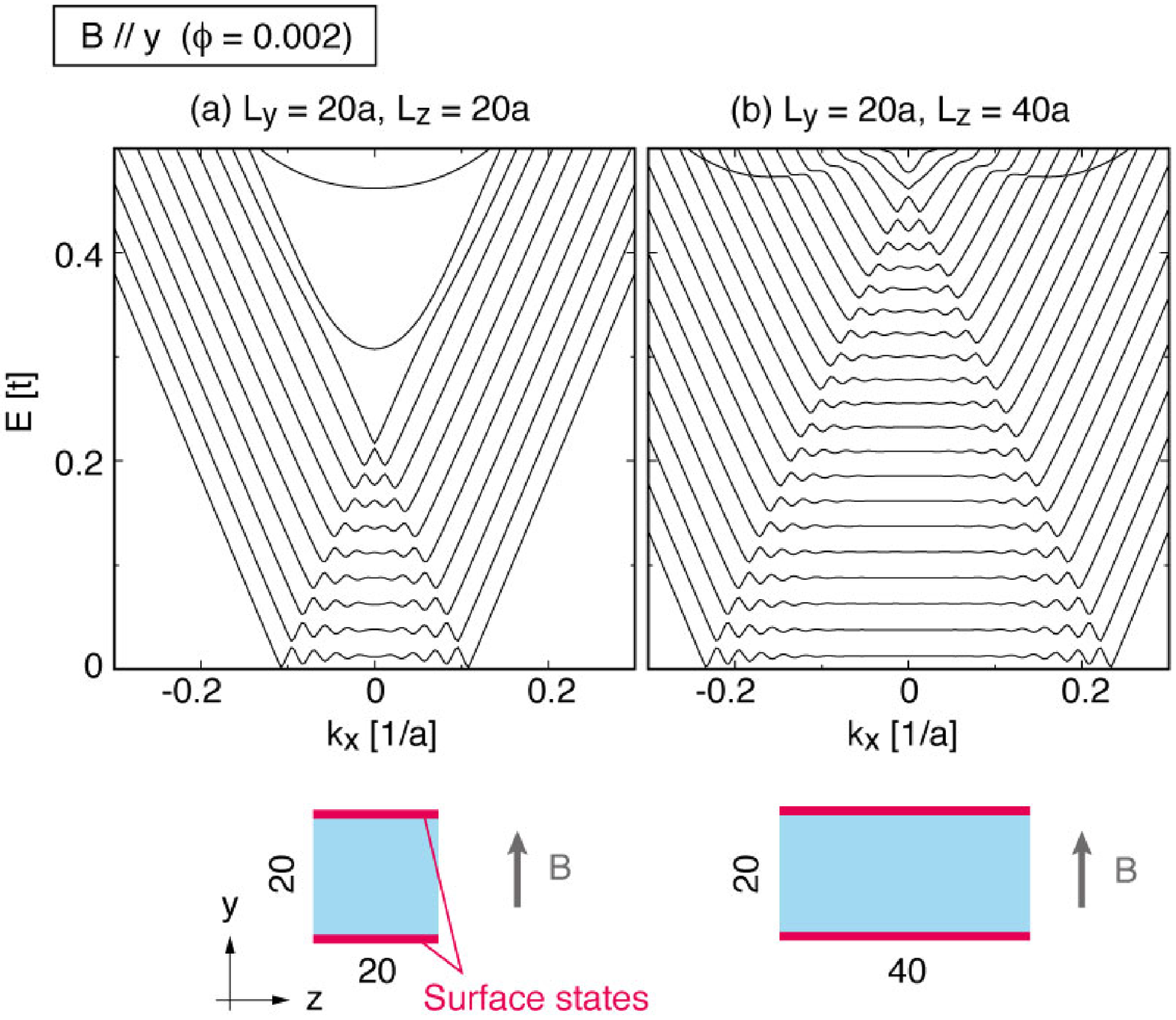}
\end{center}
\caption{Band structures calculated for different cross sections, 
(a) $(L_y,L_z) = (20a, 20a)$ and (b) $(L_y,L_z) = (20a, 40a)$,
in $B \parallel y$ with $\phi=0.002$.
}
\label{fig_band_by_aspect}
\end{figure}

\begin{figure*}
\begin{center}
\includegraphics[width=0.9\hsize]{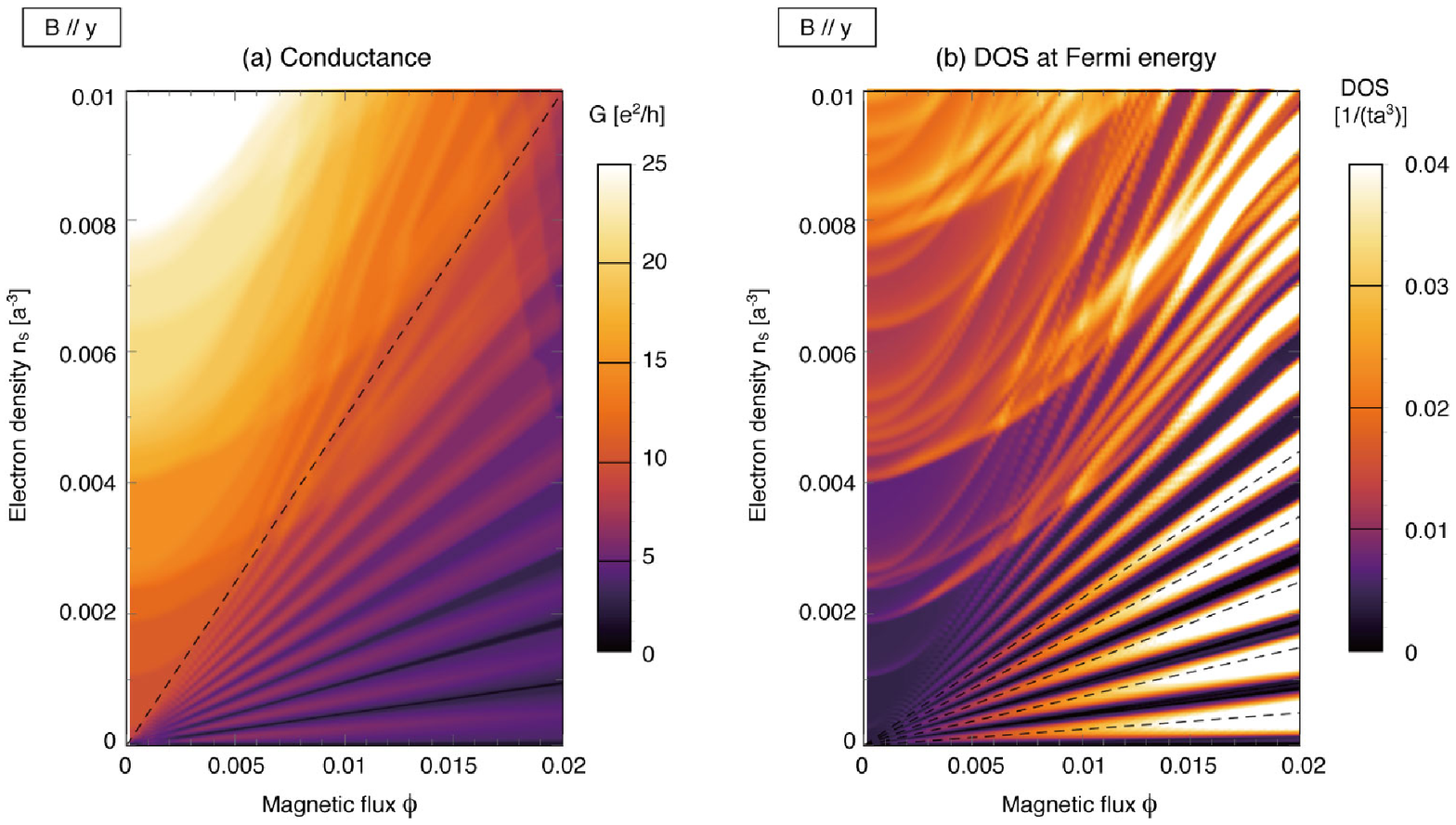}
\end{center}
\caption{Two-dimensional density maps of (a) conductance and (b) density of states 
at the Fermi level calculated for $L_y=L_z=20a$ with $B\parallel y$,
on the space of the magnetic flux $\phi$ (horizontal axis) and the carrier density $n_s$ (vertical axis).
}
\label{cond_dos_magy}
\end{figure*}

 \subsection{Perpendicular field $(B\parallel y)$} 
When a magnetic field is applied to the system,
the surface state bands are modified in different ways depending on the field direction.
We first consider $B$-field along $y$ axis,
which is perpendicular to the surface states.
Figure \ref{band_magy} shows the band structures of $L_y=L_z=20a$ 
in some different $B$-fields parallel to $y$. The amplitude of the magnetic field is specified by the flux per unit, $\phi$.
Once the field is applied, we observe that the surface bands lift the degeneracy,
where the splitting width linearly increases to $B$.
Moreover, the surface states with opposite slopes start to anticross each other,
and eventually form a series of equally-spaced flat bands in strong magnetic fields.

The splitting of the surface bands in a small $B$-field can be explained using a simple analytics as following.
At $B=0$, the surface bands labeled by different $k_z$'s are all degenerate
and they have the identical dispersion $E=\pm\hbar v k_x$.
Under the vector potential $\Vec{A} = (Bz,0,0)$,
the momentum operator in $x$ direction is shifted as
$k_x \rightarrow k_x+ eBz/\hbar$,
and then the degenerate surface states, 
which are originally plain waves in $z$-direction,
are rearranged into the Wannier-like localized states in $z$
by mixing different $k_z$'s.
These localized states have different energies 
depending on the expectation value $\langle z \rangle$, 
so that the dispersion becomes
\begin{equation}
E \approx \pm\hbar v \left( k_x + \frac{eB}{\hbar}\langle z \rangle \right).
\end{equation}
The split surface bands are schematically shown in Fig.\ \ref{band_magy}(e).
As $\langle z \rangle$ ranges from $-L_z/2$ to $L_z/2$,
the broadening width in $k_x$ direction is given by $k_b \sim eBL_z/\hbar$.
The spacing between the adjacent bands is given by 
$\Delta k =k_b / N_0 \sim 2\pi eB / (\hbar k_0)$. It gives a good estimation of the band splitting
in the numerical results.

The flat band formation in strong $B$ fields is closely related its peculiar wave function
composed of the topological surface states and bulk Landau level. 
Figure \ref{fig_wave} illustrates the wave functions 
of the lowest positive energy level at $k_x=0$ in different magnetic fields.
In $\phi=0.001$, it is nearly a linear combination of the left and right surface states,
while in stronger field $\phi=0.05$,  we see a significant wave amplitude inside the bulk
in addition to the surface amplitudes.
The bulk part actually corresponds to the $n=0$ Landau level,
of which wave function in $z$ direction is approximately given by $e^{-(z-\langle z \rangle)^2/(2l_B^2)}$,
where $l_B = \sqrt{\hbar/(eB)} = a/\sqrt{2\pi\phi}$ is the magnetic length.
The hybrid states of surface modes and Landau level were also reported in the previous works.
\cite{yang2011quantum, potter2014quantum,bulmash2016quantum}
In the present case, the center coordinate $\langle z \rangle$ is linked to $k_x$ by the relation 
$\langle z \rangle =  - k_x l_B^2$ as in the usual Landau level,
 and therefore a change in $k_x$ leads to a shift of the position in $z$.
In changing $k_x$, the energy stays constant unless $\langle z \rangle$ reaches the boundary at $\pm L_z/2$,
and that explains the flatness of the energy band within a finite region.
The $k$-space length of the flat region is therefore given by $L_z / l_B^2$, 
and it is equal to broadening width $k_b$ in Fig.\ \ref{band_magy}(e) as it should be.
The condition for the formation of well-defined flat bands is written as $L_z \gg l_B$.
The flat bands at different energies correspond to the different wave numbers in $y$ direction.
This is viewed as quantization of the $n=0$ Landau level in an infinite system, 
which linearly disperses in the momentum parallel to $B$-field.

Figure \ref{fig_band_by_aspect} presents the energy spectra for different cross sections 
at the same magnetic field $\phi=0.002$.
When comparing (a) $(L_y,L_z) = (20a, 20a)$ and (b) $(L_y,L_z) = (20a, 40a)$,
we see that the system with larger $L_z$ has a greater number of linear bands 
in accordance with Eq.\ (\ref{eq_N0}), 
and as a consequence the flat band region extends in a wider range in $k$-space.
In $L_z\to\infty$, the low-energy spectrum is dominated by flat bands.

The flat-band formation dramatically influences the electronic transport.
In Fig.\ \ref{cond_dos_magy}(a), we present the density plot of the conductance 
of the wire with $L_y=L_z=20a$ under $B\parallel y$,
on the two-dimensional space of the magnetic flux $\phi$ (horizontal axis) and the carrier density $n_s$ (vertical axis)
relative to the Weyl point.
At zero magnetic field, the conductance at low $n_s$ 
is mainly contributed from the surface channels.
If we increase $\phi$ at fixed $n_s$, i.e., move rightward on the horizontal line,
the conductance starts to dacay and oscillate after passing a certain magnetic field.
This is because the linear bands of the left-going and right-going surface modes
anticross with each other to form the flat bands,
and the number of the channels significantly decreases
leaving only a few dispersive modes near the ends of the flat bands.

The critical $B$-field at which the conductance starts to decay 
is estimated by the condition that
the diamond-shaped crossing region in Fig.\ \ref{band_magy}(e)
reaches the Fermi level, i.e., $E_b/2 > E_F$. The condition is rewritten as
\begin{equation}
B > (2\pi)^2\frac{L_y}{L_z}\frac{\hbar}{ek_0}n_s,
\label{eq_cond}
\end{equation}
or equivalently, $\phi> 2(L_y/L_z) n_s a^3$ 
 in the present model.
The criterion is indicated by the diagonal dashed line in Fig.\ \ref{cond_dos_magy}(a).
At $n_s=0$, in particular, the conductance fall takes place in an infinitesimal $B$-field
because the band anticrossing takes place right at the Fermi energy.
We also note that the slope of the dashed line against $B$ is proportional to the aspect ratio $L_z/L_y$.
In the slab geometry ($L_z/L_y\to\infty$), therefore,  the gradient of the slope becomes infinite,
so that the conductance immediately falls at any $n_s$ once $B$-field is applied.

Fig.\ \ref{cond_dos_magy}(b) shows the density map for the density of states (DOS) 
at the Fermi level.
We clearly see that the discrete flat bands give rise to an oscillatory feature,
which is consistent with the previous calculations for the slab geometry \cite{potter2014quantum,bulmash2016quantum}.
The oscillation period is determined by the number of electrons 
accomodated between the adjacent gaps in the spectrum, which is given by 
$\Delta n_s = k_b/(2\pi/L_x)/(L_xL_yL_z) = 1/(2\pi l_B^2L_y)$.
The DOS peaks are therefore specified by 
\begin{equation}
n_s \approx \frac{1}{2\pi l_B^2L_y} \left(n + \frac{1}{2}\right), \quad n=0,\pm 1,\pm 2,\cdots,
\end{equation}
which is shown as dashed lines in Fig.\ \ref{cond_dos_magy}(b).
In changing $B$-field with $n_s$ fixed, the peak appears periodically in $1/B$ as
\begin{equation}
\frac{1}{B_n} = \frac{e}{h}\frac{1}{n_s L_y} \left(n + \frac{1}{2}\right).
\label{eq_Bn}
\end{equation}
This is equivalent to the Shubnikov-de Haas oscillation of the 2D metal
with the electron area density $n_s^{\rm (2D)} = n_s L_y$.
The magnetic oscillation period Eq.\ (\ref{eq_Bn})
is independent of $k_0$, unlike the spacing in the energy axis.

If we measure the transverse voltage between $y=\pm L_y/2$, 
we should observe a Hall voltage proportional to the electric current along the wire.
Because the time-reversal symmetry is broken in the present model,
we have the Hall effect already at $B=0$, 
where the Hall conductance is proportional to the number of surface channels.
When $B$-field is applied to $y$-direction, 
the surface-mode hybridization immediately leads to a short circuit of the counter propagating channels,
and it should be observed as a sharp drop of the Hall conductance.

 \subsection{Parallel fields $(B\parallel x, z)$} 
  
\begin{figure}
\begin{center}
\leavevmode\includegraphics[width=1.\hsize]{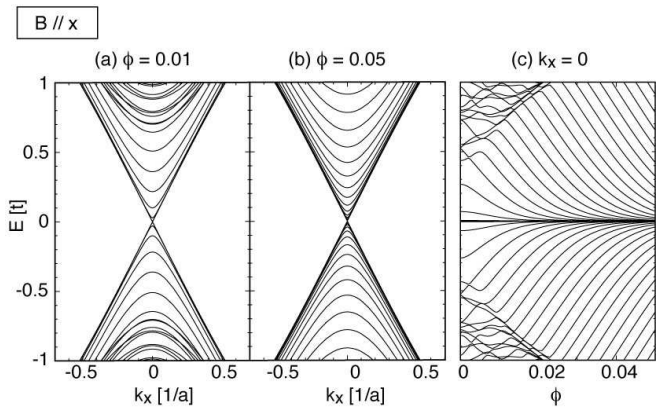}
\end{center}
\caption{Band structure of the wire of $L_y = 20 a,L_z=20a$
 in $B \parallel x$ with (a) $\phi =0.01$ (b) $\phi =0.05$. 
 Rightmost panel (c) plots the magnetic field dependence of the energy levels at $k_x=0$.
}
\label{band_magx}
\end{figure}
\begin{figure}
\begin{center}
\includegraphics[width=0.85\hsize]{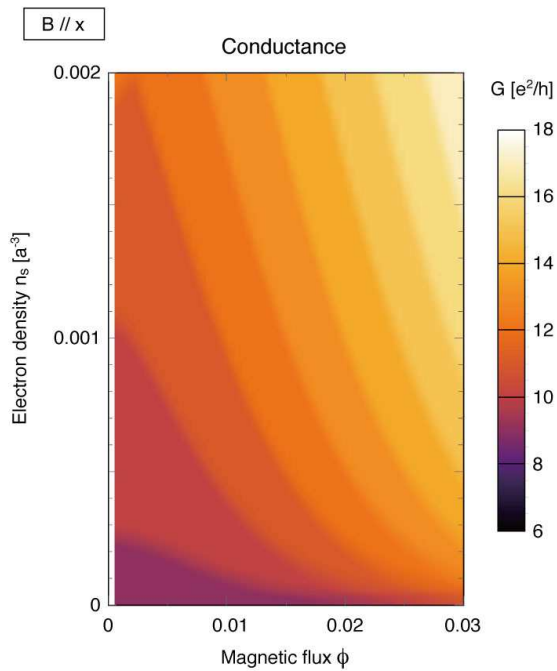}
\end{center}
\caption{Two-dimensional density maps of conductance 
in $L_y=L_z=20a$ with $B\parallel x$,
on the space of the magnetic flux $\phi$ (horizontal axis) and the carrier density $n_s$ (vertical axis).
}
\label{cond_dos_magx}
\end{figure}

The magnetic field parallel to the surface state gives rise to completely different effects.
Figures \ \ref{band_magx}(a) and (b) show the band structures of $L_y=L_z=20a$
for several $B$-fields parallel to $x$.
The rightmost panel (c) plots the magnetic field dependence of the energy levels at $k_x=0$.
In increasing $B$-field, we observe that 
the linear bands of the surface states stay almost unchanged,
while the massive bands of bulk states shift towards the zero energy,
and eventually merged into the cluster of linear bands.
This is actually understood as a result of formation of the $n=0$ Landau level:
In the magnetic field parallel to $x$ direction,
the $n=0$ level gives a linear dispersion in $k_x$
where the upslope and the downslope correspond to the Landau levels of the different Weyl nodes.
The surface state itself is barely influenced by the $B$-field in $x$ direction,
because the surface-localized wave function is insensitive to magnetic flux parallel to the surface.
Considering the extra degeneracy from the $n=0$ Landau levels,
the total number of the channels traveling in each single direction ($+x$ or $-x$) is estimated as
\begin{equation}
N \sim N_0 + \frac{L_yL_z}{2\pi l_B^2}.
\end{equation}
In increasing magnetic field, therefore, the conductance $G = (e^2/h)N$ increases one by one
every time a new subband is populated.
This property is clearly observed in the two-dimensional density map
of the conductance in Fig.\ \ref{cond_dos_magx}, which is calculated 
for $L_y=L_z=20a$ and $B\parallel x$.

In Fig.\ \ref{band_magx},
we show the band structure in $B$-fields parallel to $z$, which is another direction parallel to the surface states.
We see that the crossing point of the linear bands gradually moves upward in increasing $B$.
At $B=0$, 
the linear bands $E=\pm \hbar v k_x$ correspond to the surface states localized at $y=\pm L_y /2$, respectively.
When the $B$-field is on, 
the vector potential $A=(-By,0,0)$ shifts the wave number $k_x$ to $k_x - eBy/\hbar$,
and then the positive and negative slopes 
move horizontally in opposite direction as $E=\pm \hbar v (k_x - eB\langle y\rangle/\hbar)$
where $\langle y\rangle = \pm L_y/2$, respectively.
As a result, the crossing point moves vertically to the positive energy direction.
In Fig.\ \ref{band_magx}, the horizontal dashed (red) line indicates the Fermi energy $E_F$
at $n_s=0$.
In increasing $B$, $E_F$ goes up together with the crossing point, and finally gets into the bunch of the bulk bands.
The number of channels stays constant until that point,
after which the bulk states dominate the transport.
In increasing $\phi$, we also see that a flat region is horizontally growing in each subband.
This will eventually become a part of $n=0$ Landau level in higher magnetic field.

\begin{figure}
\begin{center}
\leavevmode\includegraphics[width=0.98\hsize]{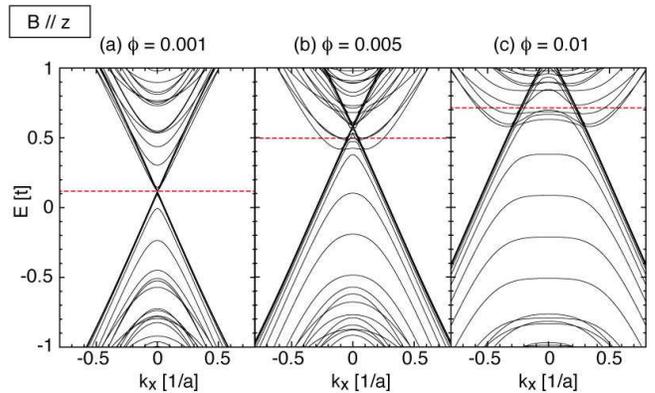}
\end{center}
\caption{Band structure of the wire of $L_y = 20 a,L_z=20a$
 in $B \parallel z$ with (a) $\phi =0.001$ (b) $\phi =0.01$ (c)$\phi =0.05$.
}

\label{band_magz}
\end{figure}

\section{Conclusion}
We studied the energy spectrum and the electric conductance of the
Weyl semimetal wire in various magnetic-field directions.
In any cases studied here, the surface states continuously change over
$n=0$ Landau levels as increasing $B$-field, but in completely different fashions 
depending on the field direction.
In particular, the field perpendicular to the surface state 
mixes up the counter-propagating surface modes into flat bands of $n=0$ Landau level.
This leads to a significant reduction of the number of the traveling modes, 
and causes an oscillatory decay of the conductance.
On the contrary, a magnetic field parallel to the wire 
form linearly-dispersed $n=0$ Landau level, and the conductance increases.

While the Fermi arc in the real Weyl semimetals is generally more complicated 
than in the present model, we expect that the conductance decay could be observed 
by choosing appropriate directions of the wire and $B$-field relative to the Fermi arc. 
The conditions for the flat-band formation, Eq.\ (\ref{eq_cond}),
depends on the Fermi arc length $k_0$ and also on the aspect ratio $L_z/L_y$
of the cross section. The critical $B$-field becomes smaller for larger $L_z/L_y$, so a thin, ribbon-like geometry 
with $L_z \gg L_y$ would be desirable to observe the effect.
Here we completely neglected the disorder effect to focus on the effects of the field-induced hybridization of the 
surface modes, so the calculation is expected to be valid in the ballistic regime
in which the wire length is shorter than the scattering mean free path.
We expect that the surface mode hybridization and the flat-band formation 
also influence the transport in the diffusive regime,
while the detailed study on the impurity effects is left for future works.

\begin{acknowledgments}
The authors thank Tetsuro Habe and Bohm-Jung Yang for helpful discussions. 
This work was supported by JSPS KAKENHI Grant Numbers JP25107001 and JP25107005.
\end{acknowledgments}

\bibliography{weylmag}

\end{document}